\documentclass[%
 aip,
 amsmath,amssymb,
 reprint,%
]{revtex4-1}

\usepackage{graphicx}
\usepackage{dcolumn}
\usepackage{bm}
\usepackage[utf8]{inputenc}
\usepackage[T1]{fontenc}
\usepackage{mathptmx}
\usepackage{etoolbox}
\usepackage{siunitx}

\makeatletter
\def\@email#1#2{%
 \endgroup
 \patchcmd{\titleblock@produce}
  {\frontmatter@RRAPformat}
  {\frontmatter@RRAPformat{\produce@RRAP{*#1\href{mailto:#2}{#2}}}\frontmatter@RRAPformat}
  {}{}
}%
\makeatother
\begin{document}

\preprint{AIP/123-QED}

\title{Resolving polarization-dependent mode dynamics in multimode fibers with 2D single-photon detector arrays}
\author{Harikumar K. Chandrasekharan*}
\email{hk47@hw.ac.uk}
\affiliation{Scottish Universities Physics Alliance, Institute of Photonics and Quantum Sciences, School of Engineering
and Physical Sciences, Heriot-Watt University, David Brewster Building, Edinburgh EH14 4AS, Scotland,
UK
}%

\author{Ross Donaldson}%
\affiliation{Scottish Universities Physics Alliance, Institute of Photonics and Quantum Sciences, School of Engineering
and Physical Sciences, Heriot-Watt University, David Brewster Building, Edinburgh EH14 4AS, Scotland,
UK
}%

\date{\today}

\begin{abstract}
Monitoring polarization dynamics in multimode fibers is critical for a range of applications, spanning from optical communication to sensing. Although the modal behavior of multimode fibers is well understood through interferometry and advanced detection techniques, most studies focus on a single polarization state of specific modes, leaving the spatial mode dynamics of the other polarization state unexplored. A variety of optical phenomena can arise during the transport of spatial modes in fibers, driven by stress-induced fiber deformations. These phenomena include changes in the effective refractive index of modes, modal dispersion, and polarization-dependent mode coupling. Observing such modal behavior typically requires complex systems, such as multi-axis interferometry. In this paper, we present quasi-real-time observation of spatial mode dynamics in a two-mode fiber using 2D single-photon avalanche diode (SPAD) arrays configured in a dual-axis setup under different fiber deformation conditions. By utilizing the time-resolved capabilities of the SPAD arrays, we capture the modal behavior of two spatial modes in the fiber during stress induced by uncontrolled deformations, enabling direct observation of spatial correlations of the modes. Additionally, we demonstrate dual-polarization mode dynamics under controlled fiber conformation, where the modal behavior remains stable within acceptable error limits. Our work opens new avenues for exploring polarization-dependent phenomena in both fundamental and applied optics, as well as in biological systems.

\end{abstract}

\maketitle

Observing polarization-dependent mode behavior in optical fibers is an area of intense research due to its applications in optical telecommunications\cite{Fatome:10, doi:10.1049}, sensing\cite{article147,article139}, and imaging\cite{article237,article576,Sivankutty:16}. Single-mode fibers (SMFs) excel in providing higher bandwidth, long-distance optical communication, and precise sensing due to the confinement of the fundamental (single) mode\cite{9678376,article222}. However, the use of SMFs is limited by higher costs and installation challenges, stemming from the delicate and precise alignment required\cite{1128517}, alongside its high tolerance to non-linear Shannon-limit\cite{Ellis:17}. On the other hand, multimode fibers (MMFs) support multiple spatial modes, due to their larger core sizes, allowing high-throughput transport of optical signal. This results in a higher potential nonlinear Shannon limit, lower costs, and simplified deployment strategies\cite{Yam:06,doi:10.1049/el:20061011}. The spatial mode profiles in MMFs are susceptible to fiber geometry, environmental perturbations, and the input polarization state. A range of optical phenomena can occur during induced perturbations, including mode mixing, modal dispersion, and interference, which significantly impact light propagation and signal integrity\cite{Schulze:15,Mathews,Qian,article7}. If harnessed effectively, such induced multimode photon dynamics can be leveraged for a broad spectrum of applications, ranging from optical communication to advanced sensing\cite{He:21,article46,DiGirolamo:07,Sarkar:24}. Characterizing these induced photon dynamics across multiple polarization states is therefore essential, especially in applications requiring high-precision signal transmission and imaging\cite{article98,article48}.      

The high-fidelity spatial information of optical signals in MMF systems can be extracted through advancements in multiplexing, mode tomography, holography, single photon detection, and machine learning techniques\cite{Cao:23,article34,Roudas:18,article21}. However, these techniques face several challenges and limitations, including mode crosstalk, computational complexity, sensitivity to alignment, and system cost, which hinder their widespread adoption. Furthermore, methods like machine learning require significant data and computational resources, while techniques such as digital holography demand precise optical alignment\cite{Zeng:21}. Innovative optical characterization techniques must be developed to overcome these challenges and advance MMF-based system performances for high-capacity communication, imaging, and sensing applications.

Regardless of the developed technology, a key capability for the future MMF-characterization system is the precise calibration of photon dynamics in two orthogonal polarization states at timescales down to several picoseconds (ps) to nanoseconds (ns). Advancements in complementary metal-oxide-semiconductor (CMOS) technology have led to the development of sophisticated 2D single-photon avalanche detector (SPAD) arrays, achieving temporal resolutions as fine as the picosecond range. With their impressive ultrafast imaging capabilities, these SPAD array detectors have proven to be ideal for various real-world applications, such as fluorescence lifetime imaging microscopy\cite{Poland:15}, time-of-flight imaging\cite{Krstajic:15,article88}, time-stretch imaging\cite{article77}, single-photon multimode imaging\cite{10.1063}, and quantum key distribution\cite{Donaldson:21}.
\begin{figure*}[htbp]
		\centering
		{\includegraphics[ width=\linewidth]{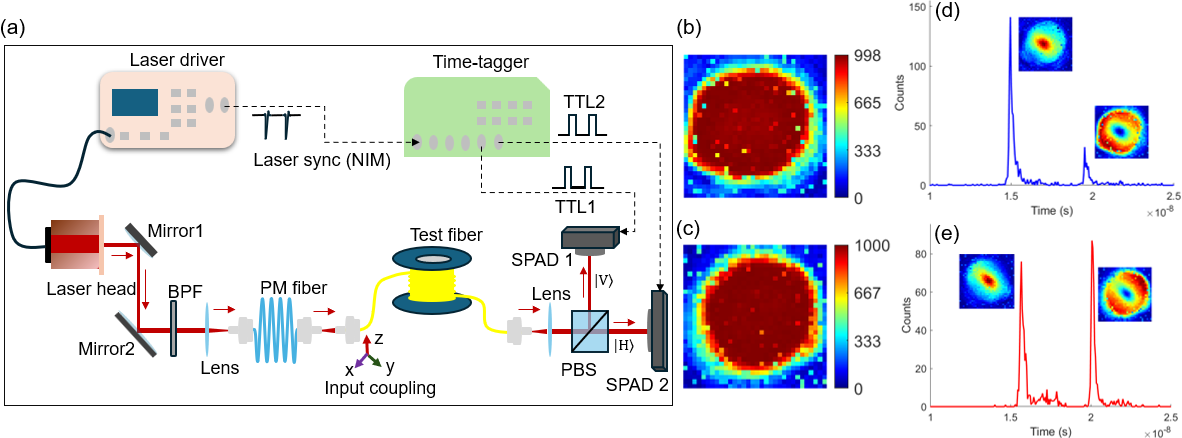}}
		\caption{(a) Experimental set-up for the dual-polarization light-in-flight spatial mode imaging system. At the two SPAD arrays, TCSPC measurement initiates when the signal from the fiber arrives at the SPAD pixels, and the periodic trigger signals from the time-tagger stops the photon measurement. (b,c) Summed counts acquired in SPAD array 1 (b) and SPAD array 2 (c)). (e,f) A single representative pixels from each SPAD array with histogram of photons. The two peaks in each histograms corresponds to LP$_{01}$ and LP$_{11}$ modes respectively. Summing the photon counts across each histograms peaks will allow the observation of mode energy distributions on two SPAD arrays as represented in the inset images.}
		\label{Fig:1}
	\end{figure*}   
  
In this work, we employ two CMOS SPAD array detectors (Photon Force Ltd) to perform quasi-real-time imaging of orthogonal polarization states in MMFs subjected to varying conformations, enabling direct observation and controlling of polarization mode dynamics. Specifically, we use two two-dimensional (2D) CMOS SPAD arrays with 32$\times$32 pixels, arranged in a square grid, forming an imaging array.  Each SPAD has a photosensitive area $\approx$ \SI{6}{\micro m} in diameter, with a pixel-pixel separation \SI{50}{\micro m}. Crucially, for our application, each pixel has a dedicated time-to-digital converter (TDC) for independent time-correlated single photon counting (TCSPC) with a dynamic range of \SI{50}{ns} and timing bin duration of 55~ps~\cite{6176268,Krstajic:15}. With our dual SPAD array setup, we were able to simultaneously capture the dynamics of two polarization states of spatial modes, we aim to elucidate the relationship between fiber geometry and polarization-dependent mode transmission characteristics. This dual-polarization imaging approach provides a comprehensive framework for investigating polarization dynamics in MMFs with high temporal resolution. 
  \begin{figure*}[tb!]
		\centering
		{\includegraphics[width=0.9\linewidth]{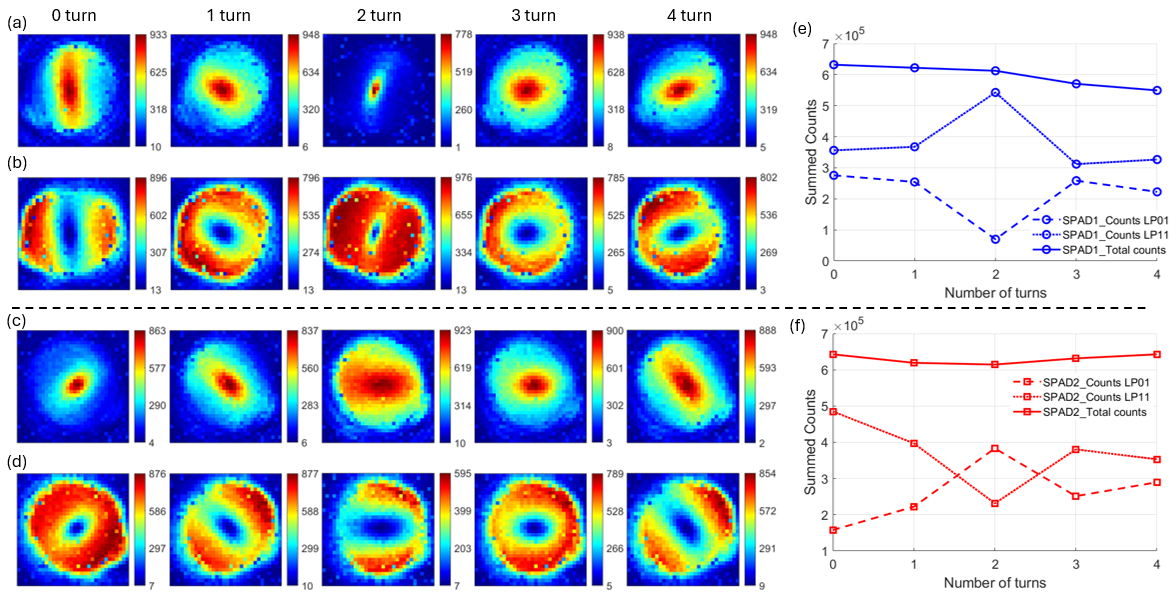}}
		\caption{Optical intensity distributions for PL$_{01}$ and LP$_{11}$ modes recorded at SPAD array 1 (a,b) and SPAD array 2 (c,d) for different fiber conformations showcasing quasi-real time dual-polarization modal dynamics. (e,f) Summed photon counts for LP$_{01}$ and LP$_{11}$ modes showcasing energy flow between modes due to mode coupling.}
	\label{fig:2}
	\end{figure*}

The experimental setup we used to observe dual-polarization mode dynamics in MMFs is shown in Figure~\ref{Fig:1}. Initially, the 2D SPAD arrays are well-calibrated in terms of instrument response function (IRF) and relative arrival time of photons using a single polarization state. For this, a pulsed laser (PicoQuant laser) with a pulse width of <70 ps at full-width at half-maximum (FWHM) was directed onto a polarization-maintaining (PM) single-mode fiber (P5-780PM-FC-2, Thorlabs) using free-space optics. The repetition rate of the laser was set to 38 MHz, at an operating wavelength of 850 nm, with a bandpass bandwidth of $\pm$ 2.5 nm. The average optical power at this rate was measured to be 2 mW at the output aperture of the laser. A bandpass filter (BPF) with a 10 nm spectral bandwidth, designed to center at 850 nm, is utilized as an angle-tunable filter at the input to narrow the laser bandwidth. After the BPF is inserted, the measured bandwidth is 0.48 nm at FWHM, with the central wavelength shifted to 852 nm. Light emitting from the distal end of the PM fiber was then collimated and projected to the SPAD array through a linear polarizer to select a single polarization state and to avoid IRF degradation in SPADs due to polarization mode dispersion. 
    
In TCSPC, photon arrival times are determined by measuring the time intervals between start and stop signals. To minimize the load on the timing electronics and achieve rapid acquisition times, the SPAD arrays operate in the "reverse start-stop" mode~\cite{Becker2005}, in which the detection of a single photon by any SPAD initiates the clock on its corresponding TDC. The clock stops upon receiving a TTL stop signal, which, in our setup, was generated by a time-tagger (ID900, ID Quantique) using the synchronizing signal (NIM) as its input. Due to the variability in the individual TDCs for each SPAD pixel, the IRF (FWHM) was found to vary from 140 ps to 388 ps for SPAD array 1, with a standard deviation of $\pm47$~ps and from 136 ps to 411 ps for SPAD array 2, with a standard deviation of $\pm42$~ps. Furthermore, because the stop signal arrives at different times to the TDCs across the SPAD pixels, the TCSPC signal peaks are aligned non-uniformly across the pixels. This shift was taken into account during the post-processing of the data, and all pixel TCSPC traces were shifted to a single pixel signal peak time for both SPAD arrays. 

After characterizing the 2D SPAD arrays in terms of peak arrival times, an experiment is conducted using a 2.4 km-long SMF-28 fiber (Corning SMF-28) as a multimode fiber (MMF)(see Figure~\ref{Fig:1}). At 852 nm, SMF-28 supports two linearly polarized (LP) spatial modes: LP$_{01}$ and LP$_{11}$. Light from the PM fiber is coupled into the proximal end of the SMF-28 via fiber-fiber butt coupling, where input coupling is optimized by adjusting the fiber facet using an XYZ-microblock stage. The distal end of the SMF-28 is imaged using a lens ($f = 3.1 mm$) and directed into a polarizing beam splitter (PBS), which separates the spatial modes into the two polarization states. Two SPAD arrays are positioned at equal optical path lengths at the image plane of the PBS outputs, with SPAD array 1 capturing the vertically polarized mode profile and SPAD array 2 capturing the horizontally polarized mode profile. Additionally, before the PBS, a portion of the SMF-28 mode profile is redirected into a CMOS camera using a non-PBS component (not shown in Fig. 1) for optical alignment and mode magnification optimization.   

The TCSPC trace of the SMF-28 mode profiles in two polarization states was recorded using SPAD arrays over a specific acquisition time. The total photon counts within the acquisition window for both detectors are presented in Figure~\ref{Fig:1} (b,c), along with a single representative pixel histogram (pixel 525) for each detector (Figure~\ref{Fig:1} (d,e)). As shown in Figure~\ref{Fig:1} (d,e), modal dispersion causes the two spatial modes to arrive at the detectors at different times, appearing as two distinct peaks. For the specific SMF-28 fiber used, the modal dispersion between the LP$_{01}$ and LP$_{11}$ modes is measured to be $\approx$ 5 ns. Summing the photon counts across the signal peaks enables precise monitoring of the mode profiles, as illustrated in the inset images. Notably, even after applying peak shift corrections to both SPAD arrays, the arrival times of modes in SPAD array 2 remain delayed by approximately 2.5 ns compared to those in SPAD array 1. This artifact is entirely electrical, primarily caused by variations in the RF cable characteristics used for sending TTL synchronization signals.

\begin{figure}[b]
		\centering
		{\includegraphics[width=\linewidth]{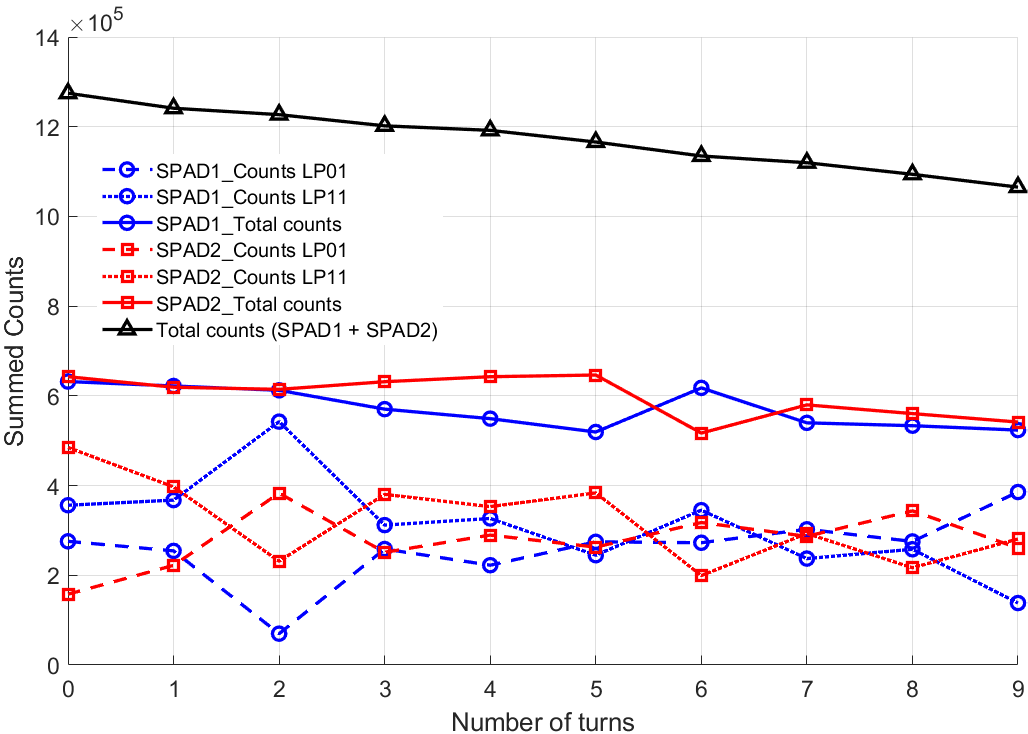}}
		\caption{Summed counts for LP$_{01}$ and LP$_{11}$ mode for two polarization states recorded for different fibre conformations showcasing quasi-real time dual-polarization mode dynamics.}
	\label{fig:3}
	\end{figure}
    
The experiment is repeated under various fiber conformations to demonstrate state-of-the-art performance. First, the effect of uncontrolled fiber conformations on mode coupling and spatial energy distribution is investigated. To this end, a 12 mm diameter pipe is used to wrap the fiber at its input end (approximately 2 m from the fiber facet). Wrapping the fiber can induce significant alterations in light propagation due to non-uniform strain distribution along different fiber locations. Previous studies have extensively investigated stress-induced changes in the refractive index and mode propagation, providing deeper insights into their effects \cite{Loterie:17, article39, Schulze:13, Schulze:15}. Figure~\ref{fig:2}(a-d) presents crucial insights into single-photon polarization-dependent mode dynamics across different fiber conformations.
    
Due to uncontrolled bending, which prevents reproducing consistent conditions, the non-uniform stress and strain on the fiber cause unpredictable changes in mode coupling and energy distribution. This effect is particularly evident in the "2-turn" case (Figure~\ref{fig:2}(a-d), column 3), where a significant discrepancy in mode energy distribution is observed between the two polarization states. To quantify these distributions, the photon counts within each LP mode are summed and plotted as a function of the number of turns, as shown in Figure~\ref{fig:2}(e,f). Notably, despite significant energy flow between modes in the "2-turn" case, the total energy remains conserved.

The experiment is then repeated for a total of 9 turns around the pipe, further quantifying mode energy distribution. As presented in Figure~\ref{fig:3}, multiple instances of unpredictable mode coupling and energy transfer are observed. Importantly, the total energy (SPAD array 1 + SPAD array 2) decreases with an increasing number of turns, a result of bend-induced loss— a well-known effect in optical fibers caused by evanescent photon leakage at bends\cite{Smink:07}.

So far, we have demonstrated dual-polarization mode dynamics, including energy flow among modes under uncontrolled fiber conformations, a critical aspect of characterizing spatially encoded modes in optical communications. This understanding is particularly relevant for mode-division multiplexing, polarization-multiplexed systems\cite{Kikuchi:11}, and fiber non-linearity management for dispersion compensation\cite{10.1063/1.5097270}. However, precise control over these mode dynamics remains challenging, as the inherent variability of manual external perturbations prevents the exact replication of identical conditions.

\begin{figure}[http]
		\centering
		{\includegraphics[width=\linewidth]{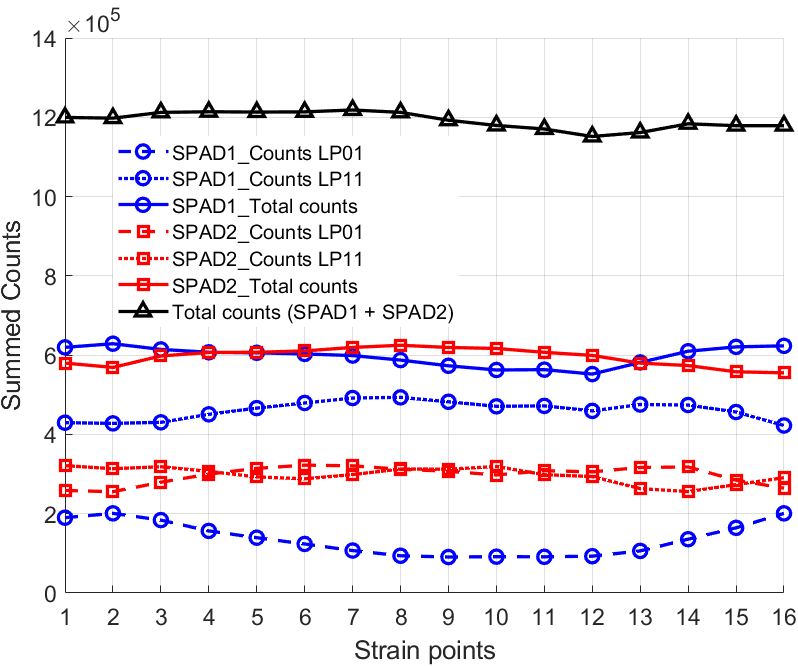}}
		\caption{Summed counts for LP$_{01}$ and LP$_{11}$ mode for two polarization states recorded for different cantilever positions showcasing quasi-real time dual-polarization mode dynamics under controlled fiber conformations.}
	\label{fig:4}
	\end{figure}

With precise control over induced deformations, the resulting mode dynamics can be harnessed for high-precision applications where the stability and reproducibility of these modal interactions are essential. To demonstrate the high degree of control over mode dynamics, we conducted a proof-of-concept experiment where fiber conformations were introduced in a more controlled manner. For this, the fiber was secured to an optical table, creating a high-strain bend along its length. A wooden cocktail stick (hereafter referred to as a cantilever) was then pressed against this high-strain section, with its position precisely controlled using an XYZ linear translation stage. We induced localized strain by systematically translating the cantilever forward along the fiber and observed its impact on mode profiles. The cantilever advanced 1 mm forward at each strain point, exerting increasing strain on the fiber. The energy flow and mode coupling across the 16 strain points are shown in Figure~\ref{fig:4}. Similar to the previous case, although inter-modal coupling induces multiple energy exchanges among modes, the total energy remains conserved.

\begin{figure*}[htbp]
		\centering
		{\includegraphics[width=\linewidth]{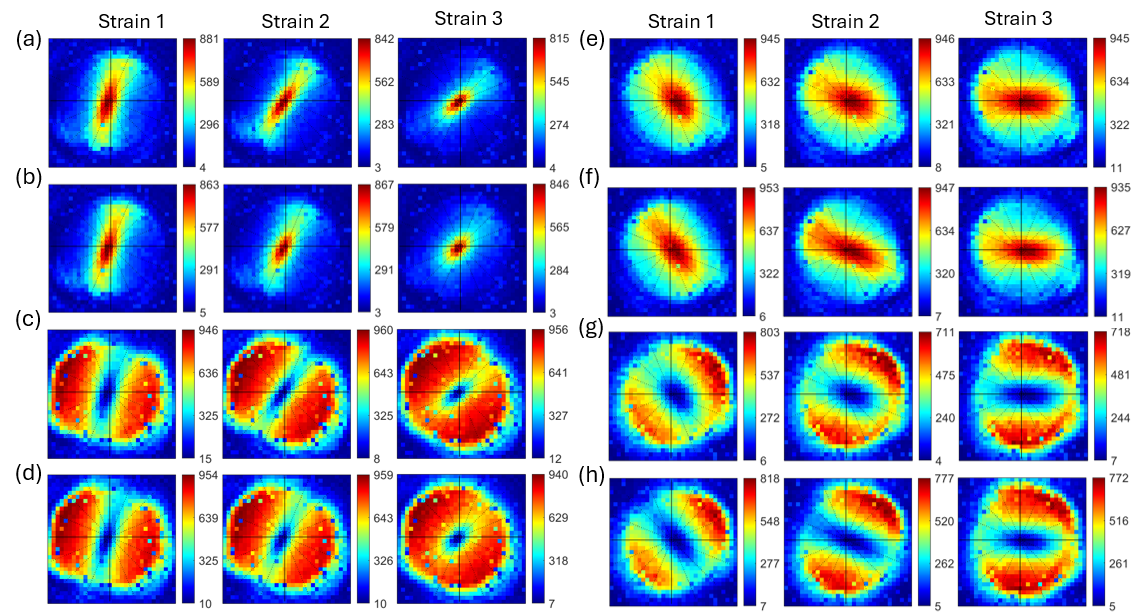}}
		\caption{Intensity distribution of LP$_{01}$ and LP$_{11}$ modes for two polarization states recorded at SPAD array 1 (a-d) and SPAD array 2 (e-h) for three different cantilever positions showcasing controlled mode dynamics and energy flow. (a,b) LP$_{01}$ mode distribution for three cantilever positions in forward (a) and reverse (b) positions for SPAD array 1. (c,d) Corresponding mode profile distribution for LP$_{11}$ mode for forward (c) and reverse (d) positions..(e,f) LP$_{01}$ mode distribution for three cantilever positions in forward (a) and reverse (b) positions for SPAD array 2. (g,h) Corresponding mode profile distributions for LP$_{11}$ mode for forward (g) and reverse (h) positions. The minor dotted lines, spaced 22.5\textdegree apart, indicate the mode orientations for comparison.}
	\label{fig:5}
	\end{figure*}

The key capability of our demonstration is that these mode orientations can be reversibly controlled by repositioning the cantilever to its respective previous positions. As presented in Figure~\ref{fig:5}, retaining the cantilever positions to the previous position results in obtaining the respective mode orientations. Negligible variations in mode orientations are observed in some profiles which are likely due to manual positioning error of the cantilever during translation stage movements, as the mode profiles are highly sensitive to small strain variations. In practical applications, automated translational stages could provide precise and repeatable strain control to generate distinct mode orientations. Remarkably, when integrated with single-pixel detectors, the demonstrated technique has the potential to replace high-data-consumption electro-optic devices. By generating well-distinguishable, multi-orientation mode profiles, it enables the creation of all-optical random patterns for high-speed compressive imaging \cite{Olivas:13,Valley:16}. Alternatively, we can generate multi-oriented mode profiles by adjusting the phase of the input light at the proximal end of the fiber\cite{article237}, though we have not included this experiment in our studies. 

Beyond sensing applications, precise strain-induced mode manipulation can enhance mode-division multiplexing by optimizing coupling efficiency. Additionally, the non-uniform refractive index variation induced by the strain can be explored to localize the mode energy across the fiber core as seen in the case of the LP$_{01}$ mode. Such controlled mode energy localization can be explored to enhance polarization-dependent coupling efficiency for optical communication, specifically free-space single-photon quantum key distribution applications, as the signal spot size significantly impacts the achievable secret key rates\cite{Lee12091}.   

In conclusion, we have demonstrated an efficient technique for observing, inducing, and calibrating dual-polarization mode dynamics in MMFs using multipixel 2D CMOS SPAD detectors. This experimental approach enables precise characterization of multimode behavior, benefiting applications such as optical communications, polarization-dependent quantum key distribution, compressive imaging, and high-precision sensing.

Beyond these areas, the dual-polarization imaging system holds significant potential for biological applications, as many biological specimens exhibit extreme sensitivity to small polarization changes due to their highly ordered structures, birefringence, or polarization-dependent scattering properties. Notable examples include collagen-rich tissues, myelinated nerve fibers, and amyloid fibrils\cite{article94,SALZER2003297,article84}.

Our approach will open a route to polarization-sensitive MMF characterization technologies without complicated optical systems or computational requirements. With further optimization, the technique can be extended to fibers supporting a larger number of spatial modes, broadening its applicability to areas such as polarization-dependent quantum entanglement\cite{10.1063888,Jones2018TuningQC} and beyond.
\newline

\vspace*{3mm}
\noindent{\bf ACKNOWLEDGMENTS}\\
The authors thank Photon Force Limited, UK for providing the SPAD array detectors used in this work. The authors thank Innovate UK (10002685); Royal Academy of Engineering RF\textbackslash201718\textbackslash1746); Engineering and Physical Sciences Research Council (EP/T001011/1) for funding support.
\newline\\
\noindent{\bf AUTHOR DECLARATIONS}\\

\noindent{\bf Conflict of Interest}\\
The authors declare no conflict of interests.

\noindent{\bf Author Contributions}\\
HKC proposed the dual-polarization multimode dynamics imaging concept, characterized the experimental setup, and performed the measurements. HKC analyzed the data, discussed experiments and finalized the results with RD. The manuscript was written by HKC and later reviewed by RD. Funding acquisition and project administration by RD. 

\vspace*{2mm}
\noindent{\bf DATA AVAILABILITY}\\
The data that support the findings of this study are openly available in the Heriot-Watt University PURE research data management system.
\newline\\
\noindent{\bf References}
\bibliographystyle{unsrt}

\begin{thebibliography}{10}

\bibitem{Fatome:10}
J.~Fatome, S.~Pitois, P.~Morin, and G.~Millot.
\newblock Observation of light-by-light polarization control and stabilization in optical fibre for telecommunication applications.
\newblock {\em Opt. Express}, 18(15):15311--15317, Jul 2010.

\bibitem{doi:10.1049}
G.B. Xavier, T.R. da~Silva, G.P. Temporão, and J.P. von~der Weid.
\newblock Polarisation drift compensation in 8 km-long {Mach-Zehnder} fibre-optical interferometer for quantum communication.
\newblock {\em Electronics Letters}, 47:608--609, 2011.

\bibitem{article147}
Fangcheng Shen, Kaiming Zhou, Changle Wang, Haiming Jiang, Di~Peng, Hongyan Xia, Kang Xie, and Lin Zhang.
\newblock Polarization dependent cladding modes coupling and spectral analyses of excessively tilted fiber grating.
\newblock {\em Optics Express}, 28, 01 2020.

\bibitem{article139}
Jose Ferrari, Erna Frins, and Wolfgang Dultz.
\newblock Optical fiber vibration sensor using ({P}ancharatnam) phase step interferometry.
\newblock {\em Journal of Lightwave Technology}, 15:968 -- 971, 07 1997.

\bibitem{article237}
Wen Xiong, Chia~Wei Hsu, Yaron Bromberg, Jose Antonio-Lopez, Rodrigo Correa, and Hui Cao.
\newblock Complete polarization control in multimode fibers with polarization and mode coupling.
\newblock {\em Light: Science \& Applications}, 7, 09 2017.

\bibitem{article576}
Runze Zhu, Haogong Feng, and Fei Xu.
\newblock Deep learning-based multimode fiber imaging in multispectral and multipolarimetric channels.
\newblock {\em Optics and Lasers in Engineering}, 161:107386, 02 2023.

\bibitem{Sivankutty:16}
Siddharth Sivankutty, Esben~Ravn Andresen, G\'{e}raud Bouwmans, Thomas~G. Brown, Miguel~A. Alonso, and Herv\'{e} Rigneault.
\newblock Single-shot polarimetry imaging of multicore fiber.
\newblock {\em Opt. Lett.}, 41(9):2105--2108, May 2016.

\bibitem{9678376}
Fabio Pittalà, Ralf-Peter Braun, Georg Böcherer, Patrick Schulte, Maximilian Schaedler, Stefano Bettelli, Stefano Calabrò, Maxim Kuschnerov, Andreas Gladisch, Fritz-Joachim Westphal, Changsong Xie, Rongfu Chen, Qibing Wang, and Bofang Zheng.
\newblock 1.71 {Tb/s} single-channel and 56.51 {Tb/s} {DWDM} transmission over 96.5 km field-deployed {SSMF}.
\newblock {\em IEEE Photonics Technology Letters}, 34(3):157--160, 2022.

\bibitem{article222}
Bill Corcoran, Mengxi Tan, Xingyuan Xu, Andreas Boes, Jiayang Wu, T.~Nguyen, Sai Chu, Brent Little, and David Moss.
\newblock Ultra-dense optical data transmission over standard fibre with a single chip source.
\newblock {\em Nature Communications}, 11, 05 2020.

\bibitem{1128517}
A.~Cardama and E.T. Kornhauser.
\newblock Modal analysis of coupling problems in optical fibers.
\newblock {\em IEEE Transactions on Microwave Theory and Techniques}, 23(1):162--169, 1975.

\bibitem{Ellis:17}
A.~D. Ellis, M.~E. McCarthy, M.~A. Z.~Al Khateeb, M.~Sorokina, and N.~J. Doran.
\newblock Performance limits in optical communications due to fiber nonlinearity.
\newblock {\em Adv. Opt. Photon.}, 9(3):429--503, Sep 2017.

\bibitem{Yam:06}
Scott S.-H. Yam and Frank Achten.
\newblock High-speed data transmission over a 1 km broad wavelength window multimode fiber.
\newblock {\em Opt. Lett.}, 31(13):1954--1956, Jul 2006.

\bibitem{doi:10.1049/el:20061011}
S.S.-H. Yam and F.~Achten.
\newblock Single wavelength 40 {Gbit/s} transmission over 3.4 km broad wavelength window multimode fibre.
\newblock {\em Electronics Letters}, 42:592--594, 2006.

\bibitem{Schulze:15}
Christian Schulze, Robert Br\"{u}ning, Siegmund Schr\"{o}ter, and Michael Duparr\'{e}.
\newblock Mode coupling in few-mode fibers induced by mechanical stress.
\newblock {\em J. Lightwave Technol.}, 33(21):4488--4496, Nov 2015.

\bibitem{Mathews}
Maxime Matthès, Yaron Bromberg, Julien Rosny, and Sébastien Popoff.
\newblock Learning and avoiding disorder in multimode fibers.
\newblock {\em Physical Review X}, 11, 06 2021.

\bibitem{Qian}
Sen Qian, Yang Xu, Lisheng Zhong, and Lei Su.
\newblock Power flow in a large-core multimode fiber under external perturbation and its applications.
\newblock {\em Scientific Reports}, 7, 12 2017.

\bibitem{article7}
Ronen Shekel, Ohad Lib, Rodrigo Gutiérrez-Cuevas, Sébastien Popoff, Alexander Ling, and Yaron Bromberg.
\newblock Shaping single photons through multimode optical fibers using mechanical perturbations.
\newblock {\em APL Photonics}, 8, 09 2023.

\bibitem{He:21}
Weitao He, Ruihuan Wu, Weiyi Hong, and Aiping Luo.
\newblock Modal dynamics in multimode optical fibers: an attractor of high-order modes.
\newblock {\em Opt. Express}, 29(20):32682--32690, Sep 2021.

\bibitem{article46}
Yiyu Zhou, Boris Braverman, Alexander Fyffe, Runzhou Zhang, Jiapeng Zhao, Alan Willner, Zhimin Shi, and Robert Boyd.
\newblock High-fidelity spatial mode transmission through a 1-km-long multimode fiber via vectorial time reversal.
\newblock {\em Nature Communications}, 12, 03 2021.

\bibitem{DiGirolamo:07}
Salvatore~Di Girolamo, Alexei~A. Kamshilin, Roman~V. Romashko, Yuriy~N. Kulchin, and Jean~C. Launay.
\newblock Sensing of multimode-fiber strain by a dynamic photorefractive hologram.
\newblock {\em Opt. Lett.}, 32(13):1821--1823, Jul 2007.

\bibitem{Sarkar:24}
Mohammad Nazmul~Islam Sarkar, Linh~Viet Nguyen, Adam~D. Kilpatrick, David~G. Lancaster, and Stephen~C. Warren-Smith.
\newblock Dynamic multimode fiber specklegram sensor smart bed enabled by deep learning.
\newblock {\em J. Lightwave Technol.}, 42(18):6342--6350, Sep 2024.

\bibitem{article98}
Piergiorgio Caramazza, Oisín Moran, Roderick Murray-Smith, and Daniele Faccio.
\newblock Transmission of natural scene images through a multimode fibre.
\newblock {\em Nature Communications}, 10, 05 2019.

\bibitem{article48}
Haoyi Yu, Zihao Huang, Simone Lamon, Baokai Wang, Haibo Ding, Jian Lin, Qi~Wang, Haitao Luan, Min Gu, and Qiming Zhang.
\newblock All-optical image transportation through a multimode fibre using a miniaturized diffractive neural network on the distal facet.
\newblock {\em Nature Photonics}, pages 1--8, 02 2025.

\bibitem{Cao:23}
Hui Cao, Tom\'{a}\v{s} \v{C}i\v{z}m\'{a}r, Sergey Turtaev, Tom\'{a}\v{s} Tyc, and Stefan Rotter.
\newblock Controlling light propagation in multimode fibers for imaging, spectroscopy, and beyond.
\newblock {\em Adv. Opt. Photon.}, 15(2):524--612, Jun 2023.

\bibitem{article34}
Yuanhang Zhang, Nicolas Fontaine, Mikael Mazur, Haoshuo Chen, Roland Ryf, Guifang Li, and Andrea Blanco-Redondo.
\newblock Impulse response characterization of a commercial multimode fiber using superconducting nanowire single-photon detectors.
\newblock {\em Journal of Lightwave Technology}, 40:1--1, 08 2022.

\bibitem{Roudas:18}
Ioannis Roudas, Jaroslaw Kwapisz, and Daniel~A. Nolan.
\newblock Optimal launch states for the measurement of principal modes in optical fibers.
\newblock {\em J. Lightwave Technol.}, 36(20):4915--4931, Oct 2018.

\bibitem{article21}
Stefan Rothe, Qian Zhang, Nektarios Koukourakis, and Jürgen Czarske.
\newblock Intensity-only mode decomposition on multimode fibers using a densely connected convolutional network.
\newblock {\em Journal of Lightwave Technology}, PP:1--1, 11 2020.

\bibitem{Zeng:21}
Tianjiao Zeng, Yanmin Zhu, and Edmund~Y. Lam.
\newblock Deep learning for digital holography: a review.
\newblock {\em Opt. Express}, 29(24):40572--40593, Nov 2021.

\bibitem{Poland:15}
Simon~P. Poland, Nikola Krstaji\'{c}, James Monypenny, Simao Coelho, David Tyndall, Richard~J. Walker, Viviane Devauges, Justin Richardson, Neale Dutton, Paul Barber, David Day-Uei Li, Klaus Suhling, Tony Ng, Robert~K. Henderson, and Simon~M. Ameer-Beg.
\newblock A high speed multifocal multiphoton fluorescence lifetime imaging microscope for live-cell fret imaging.
\newblock {\em Biomed. Opt. Express}, 6(2):277--296, Feb 2015.

\bibitem{Krstajic:15}
Nikola Krstaji\'{c}, Simon Poland, James Levitt, Richard Walker, Ahmet Erdogan, Simon Ameer-Beg, and Robert~K. Henderson.
\newblock {0.5 billion events per second time correlated single photon counting using CMOS SPAD arrays}.
\newblock {\em Opt. Lett.}, 40(18):4305--4308, 2015.

\bibitem{article88}
Genevieve Gariepy, Nikola Krstajić, Robert Henderson, Chunyong Li, Robert Thomson, Gerald Buller, Barmak Heshmat, Ramesh Raskar, Jonathan Leach, and Daniele Faccio.
\newblock Single-photon sensitive light-in-flight imaging.
\newblock {\em Nature communications}, 6:6021, 02 2015.

\bibitem{article77}
Harikumar K~Chandrasekharan, Frauke Izdebski, Itandehui Gris~Sánchez, Nikola Krstajić, Richard Walker, Helen Bridle, Paul Dalgarno, William Macpherson, Robert Henderson, Tim Birks, and Robert Thomson.
\newblock Multiplexed single-mode wavelength-to-time mapping of multimode light.
\newblock {\em Nature Communications}, 8, 04 2016.

\bibitem{10.1063}
Harikumar~K. Chandrasekharan, Katjana Ehrlich, Michael~G. Tanner, Dionne~M. Haynes, Sebabrata Mukherjee, Tim~A. Birks, and Robert~R. Thomson.
\newblock Observing mode-dependent wavelength-to-time mapping in few-mode fibers using a single-photon detector array.
\newblock {\em APL Photonics}, 5(6):061303, 06 2020.

\bibitem{Donaldson:21}
Ross Donaldson, Dmytro Kundys, Aurora Maccarone, Robert Henderson, Gerald~S. Buller, and Alessandro Fedrizzi.
\newblock Towards combined quantum bit detection and spatial tracking using an arrayed single-photon sensor.
\newblock {\em Opt. Express}, 29(6):8181--8198, Mar 2021.

\bibitem{6176268}
M.~{Gersbach}, Y.~{Maruyama}, R.~{Trimananda}, M.~W. {Fishburn}, D.~{Stoppa}, J.~A. {Richardson}, R.~{Walker}, R.~{Henderson}, and E.~{Charbon}.
\newblock {A Time-Resolved, Low-Noise Single-Photon Image Sensor Fabricated in Deep-Submicron CMOS Technology}.
\newblock {\em IEEE Journal of Solid-State Circuits}, 47(6):1394--1407, 2012.

\bibitem{Becker2005}
W.~Becker.
\newblock Advanced time- correlated single photon counting techniques.
\newblock Springer, Berlin, Heidelberg, 2005.

\bibitem{Loterie:17}
Damien Loterie, Demetri Psaltis, and Christophe Moser.
\newblock Bend translation in multimode fiber imaging.
\newblock {\em Opt. Express}, 25(6):6263--6273, Mar 2017.

\bibitem{article39}
Martin Plöschner, Tomáš Tyc, and Tomáš Čižmár.
\newblock Seeing through chaos in multimode fibres.
\newblock {\em Nat Photon}, 9:529--535, 08 2015.

\bibitem{Schulze:13}
Christian Schulze, Adrian Lorenz, Daniel Flamm, Alexander Hartung, Siegmund Schr\"{o}ter, Hartmut Bartelt, and Michael Duparr\'{e}.
\newblock Mode resolved bend loss in few-mode optical fibers.
\newblock {\em Opt. Express}, 21(3):3170--3181, Feb 2013.

\bibitem{Smink:07}
Rutger~W. Smink, Bastiaan~P. de~Hon, and Anton~G. Tijhuis.
\newblock Bending loss in optical fibers---a full-wave approach.
\newblock {\em J. Opt. Soc. Am. B}, 24(10):2610--2618, Oct 2007.

\bibitem{Kikuchi:11}
Kazuro Kikuchi.
\newblock Analyses of wavelength- and polarization-division multiplexed transmission characteristics of optical quadrature-amplitude-modulation signals.
\newblock {\em Opt. Express}, 19(19):17985--17995, Sep 2011.

\bibitem{10.1063/1.5097270}
Lothar Moeller.
\newblock Nonlinear depolarization of light in optical communication fiber.
\newblock {\em APL Photonics}, 5(5):050801, 05 2020.

\bibitem{Olivas:13}
Stephen~J. Olivas, Yaron Rachlin, Lydia Gu, Brian Gardiner, Robin Dawson, Juha-Pekka Laine, and Joseph~E. Ford.
\newblock Characterization of a compressive imaging system using laboratory and natural light scenes.
\newblock {\em Appl. Opt.}, 52(19):4515--4526, Jul 2013.

\bibitem{Valley:16}
George~C. Valley, George~A. Sefler, and T.~Justin Shaw.
\newblock Multimode waveguide speckle patterns for compressive sensing.
\newblock {\em Opt. Lett.}, 41(11):2529--2532, Jun 2016.

\bibitem{Lee12091}
Alexandra Lee, Alfonso~Tello Castillo, Craig Whitehill, and Ross Donaldson.
\newblock The impact of spot-size on single-photon avalanche diode timing-jitter and quantum key distribution.
\newblock {\em IET Quantum Communication}, 1-7(7), 2024.

\bibitem{article94}
Chao He, Honghui He, Jintao Chang, Binguo Chen, Hui Ma, and Martin Booth.
\newblock Polarisation optics for biomedical and clinical applications: a review.
\newblock {\em Light: Science \& Applications}, 10:194, 09 2021.

\bibitem{SALZER2003297}
James~L Salzer.
\newblock Polarized domains of myelinated axons.
\newblock {\em Neuron}, 40(2):297--318, 2003.

\bibitem{article84}
Haitham Shaban, Cesar Valades, Julien Savatier, and Sophie Brasselet.
\newblock Polarized super-resolution structural imaging inside amyloid fibrils using thioflavine t.
\newblock {\em Scientific Reports}, 7, 10 2017.

\bibitem{10.1063888}
Yicheng Shi, Soe Moe~Thar, Hou~Shun Poh, James~A. Grieve, Christian Kurtsiefer, and Alexander Ling.
\newblock Stable polarization entanglement based quantum key distribution over a deployed metropolitan fiber.
\newblock {\em Applied Physics Letters}, 117(12):124002, 09 2020.

\bibitem{Jones2018TuningQC}
Daniel~E. Jones, Brian~T. Kirby, and Michael Brodsky.
\newblock Tuning quantum channels to maximize polarization entanglement for telecom photon pairs.
\newblock {\em npj Quantum Information}, 4:1--7, 2018.

\end{thebibliography}
\providecommand{\noopsort}[1]{}\providecommand{\singleletter}[1]{#1}%

\end{document}